\documentclass{PoS}

\title{Status of the DeeMe Experiment,\\
  an Experimental Search for $\mu$-$e$ Conversion\\
  at J-PARC MLF}

\ShortTitle{Status of an Experimental Search for $\mu$-$e$ Conversion at J-PARC MLF}

\author{\speaker{N. Teshima}\\
        Graduate School of Science, Osaka City University\\
        Nambu Yoichiro Institute of Theoretical and Experimental Physics\\
        E-mail: \email{teshima@ocupc1.hep.osaka-cu.ac.jp}}

\abstract{The DeeMe experiment is planned to search for muon-to-electron conversion at J-PARC MLF. Our goal is to measure the process with a single event sensitivity of $1 \times 10^{-13}$ or $2 \times 10^{-14}$ for a graphite or silicon carbide target. That is one or two orders of magnitude better than the current upper limits, $7 \times 10^{-13}$ for a gold target by the SINDRUM-II experiment at PSI and $4.6 \times 10^{-12}$ for a titanium target by the experiment at TRIUMF. We are in the middle of preparation for the experiment. The construction of the secondary beamline, H Line, is now in progress. Four tracking detectors have been manufactured in 2017, and the optimization study of the filling gas is ongoing to improve the performance. After getting the better hit efficiency, we measured the momentum spectrum of electrons produced through muon-decay-in-orbit from a carbon target at J-PARC MLF D2 Area in March, 2019. In this paper, the preparation status of the DeeMe experiment will be presented.}

\FullConference{The 21st international workshop on neutrinos from accelerators (NuFact2019)\\
		August 26 - August 31, 2019\\
		Daegu, Korea}

\usepackage{url}
\usepackage{mathrsfs}

\begin{document}

\section*{The Collaboration}
\noindent Collaborators as of November in 2019.\\[7pt]
M. Aoki$^{(1)}$, D. Bryman$^{(2)}$, Y. Higashino$^{(1)}$, M. Ikegami$^{(3)}$, H. Ikeuchi$^{(7)}$, Y. Irie$^{(3)}$, S. Ito$^{(12)}$,\\
N. Kawamura$^{(3)}$, M. Kinsho$^{(4)}$, K. Komukai$^{(7)}$, S. Makimura$^{(3)}$, S. Meigo$^{(4)}$, T. Mibe$^{(3)}$,\\
S. Mihara$^{(3)}$, Y. Miyake$^{(3)}$, D. Nagao$^{(1)}$, Y. Nakatsugawa$^{(5)}$, H. Natori$^{(3)}$, H. Nishiguchi$^{(3)}$,\\
T. Numao$^{(6)}$, C. Ohomori$^{(3)}$, P. K. Saha$^{(4)}$, N. Saito$^{(3)}$, Y. Seiya$^{(7,8)}$, K. Shigaki$^{(9)}$,\\
K. Shimomura$^{(3)}$, P. Strasser$^{(3)}$, T. Takahashi$^{(7)}$, N. Teshima$^{(7,8)}$, N. D. Thong$^{(10)}$,\\
N. M. Truong$^{(11)}$, K. Yamamoto$^{(3)}$, K. Yamamoto$^{(7,8)}$, S. Yamashina$^{(1)}$, T. Yamazaki$^{(3)}$,\\
M. Yoshii$^{(3)}$, K. Yoshimura$^{(12)}$\\[7pt]
(1) Osaka University, (2) UBC, (3) KEK, (4) JAEA, (5) IHEP, (6) TRIUMF,\\
(7) Osaka City University, (8) NITEP, (9) Hiroshima University,\\
(10) Vietnam National University Ho Chi Minh City, (11) University of California, Davis,\\
(12) Okayama University

\section{Introduction}
\subsection{Charged Lepton Flavor Violation and New Physics}
\noindent In the Standard Model, charged lepton flavor violating processes are forbidden, and we have not yet observed those processes. If we take into account neutrino oscillations, the branching ratio for $\mu \rightarrow e \gamma$ is suppressed below to $10^{-54}$. That is too low probability to observe it in the experiments. However, some theoretical models beyond the Standard Model predict observable branching ratios $10^{-13}$ to $10^{-17}$. An observation of charged lepton flavor violating processes at large rate thus means the existence of new physics.

\subsection{Photonic and Non-Photonic Processes}
\noindent Possible processes for charged lepton flavor violation can be classified as photonic and non-photonic, and these can be expressed by an effective Lagrangian below \cite{photonic}:
\begin{eqnarray}
\mathcal{L} = \frac{1}{1+\kappa}\frac{m_{\mu}}{\Lambda^{2}}
\overline{\mu_{R}}\sigma^{\mu\nu}e_{L}F_{\mu\nu} + \frac{\kappa}{1+\kappa}\frac{1}{\Lambda^{2}}(\overline{\mu_{L}}\gamma^{\mu}e_{L})(\overline{q_{L}}\gamma_{\mu}q_{L})
\end{eqnarray}
where $\kappa$ is the relative strength of photonic and non-photonic processes, and $\Lambda$ is the mass scale of new physics. For pure photonic processes, the branching ratio for $\mu$-$e$ conversion will be approximately one hundredth of that for $\mu \rightarrow e \gamma$. On the other hand, if the non-photonic contribution dominates, only $\mu$-$e$ conversion will occur. It is important to probe the charged lepton flavor violation with as many different approaches as possible.

\section{DeeMe Experiment}
\begin{figure}[htbp]
\begin{minipage}[t]{0.29\hsize}
  \centering
  \includegraphics[width=\hsize]{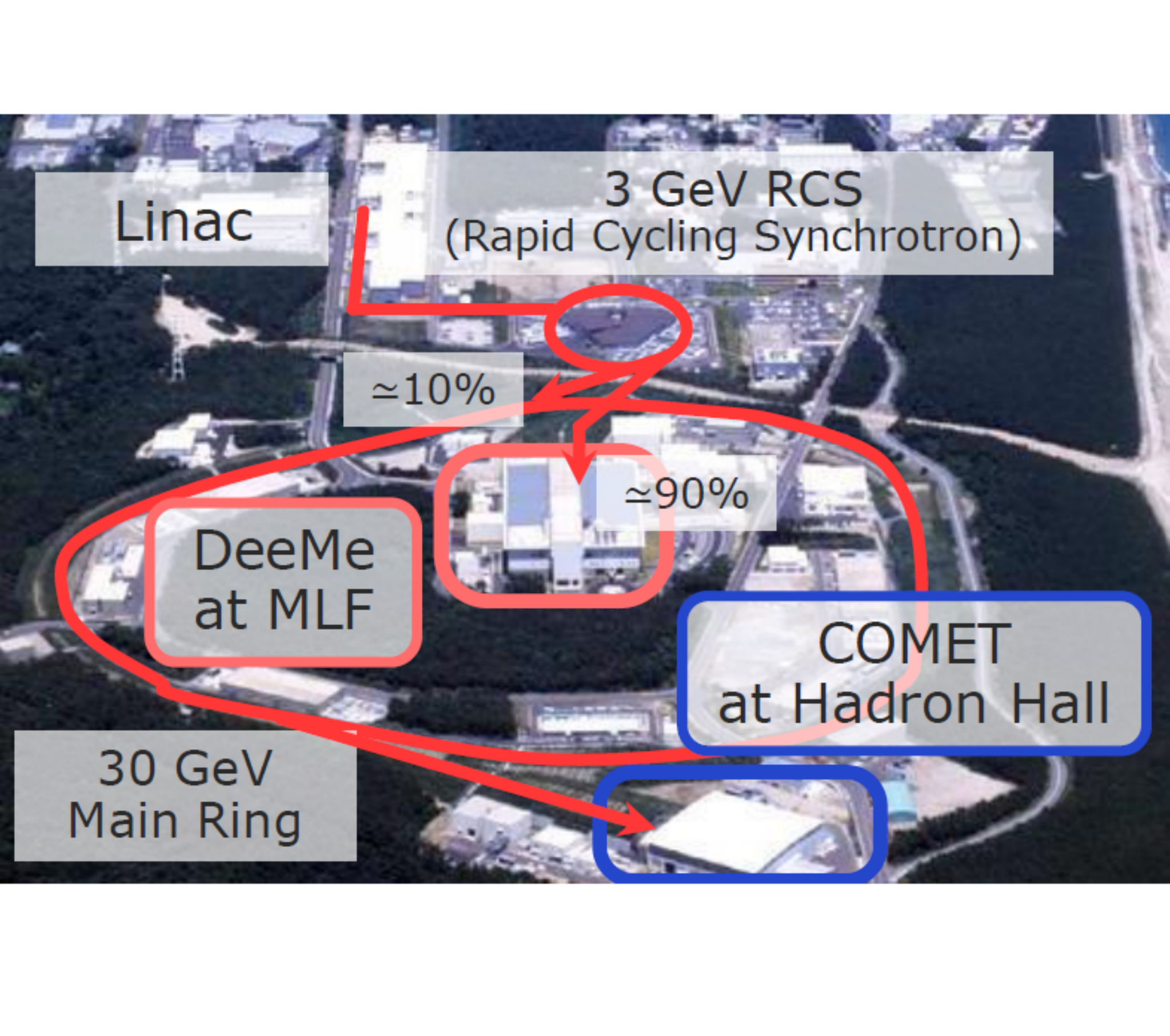}
  \caption{Photo of an overall perspective of J-PARC.}
  \label{fig:jparc}
\end{minipage}
\begin{minipage}[t]{0.05\hsize}
\hspace{0.05\hsize}
\end{minipage}
\begin{minipage}[t]{0.29\hsize}
  \centering
  \includegraphics[width=\hsize]{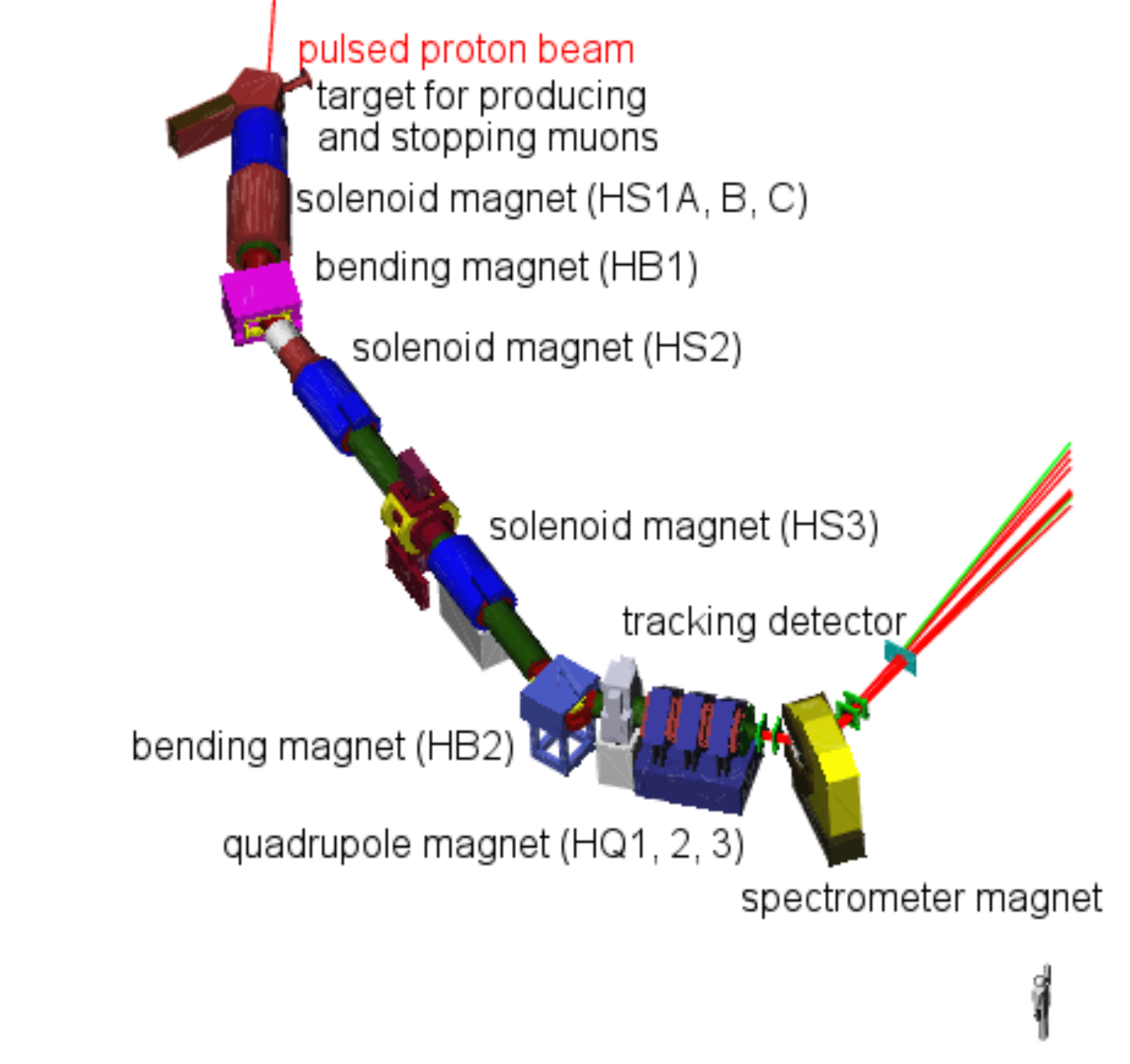}
  \caption{Experimental apparatus drawn in G4beamline simulation.}
  \label{fig:hline2}
\end{minipage}
\begin{minipage}[t]{0.05\hsize}
\hspace{0.05\hsize}
\end{minipage}
\begin{minipage}[t]{0.28\hsize}
  \centering
  \includegraphics[width=\hsize]{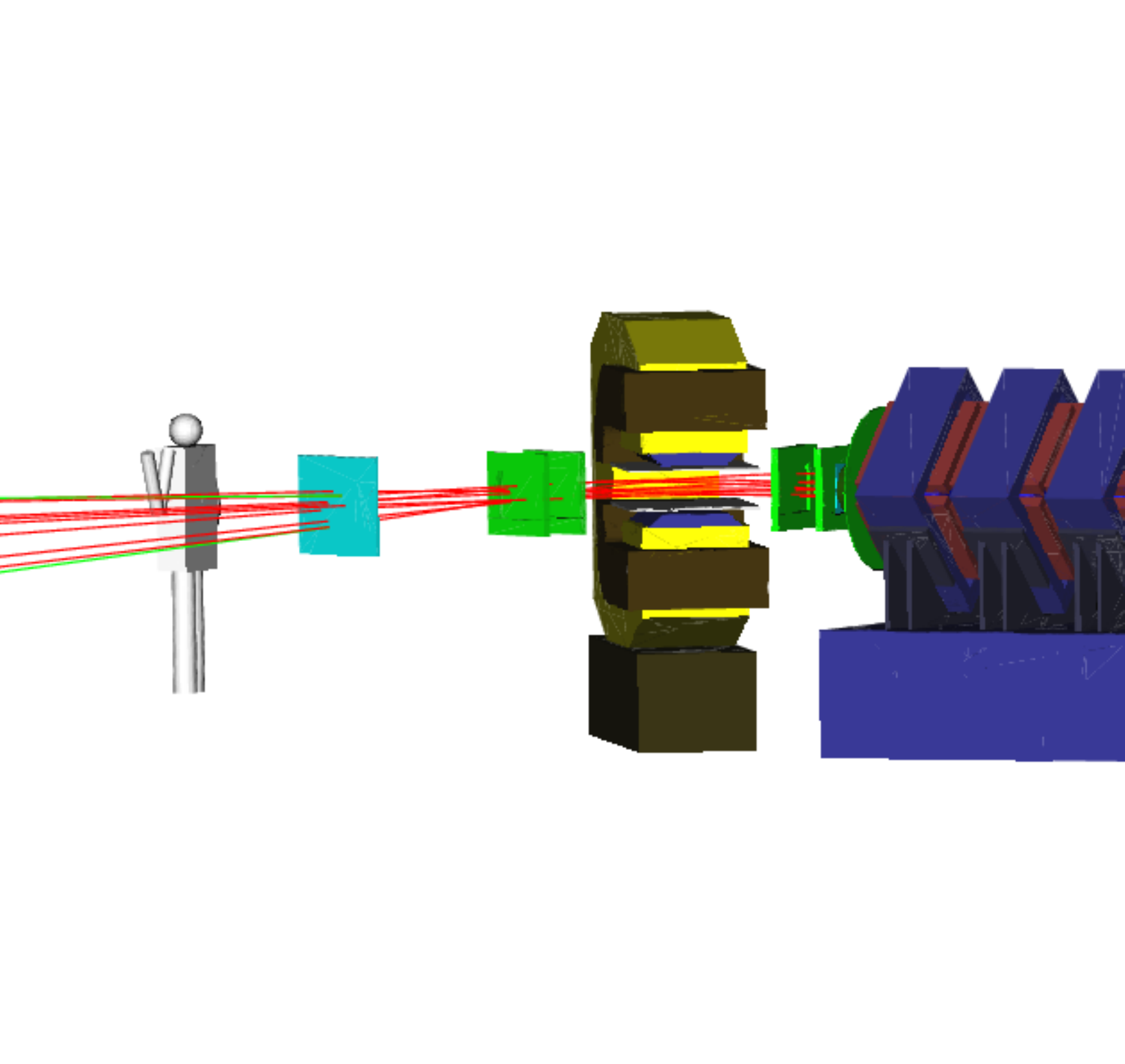}
  \caption{Enlarged view of the spectrometer in Fig.~\ref{fig:hline2}.}
  \label{fig:hline1}
\end{minipage}
\end{figure}
\subsection{Experimental Facility}
\noindent The DeeMe Experiment is planned to search for $\mu$-$e$ conversion at J-PARC Material and Life Science Experimental Facility (MLF) in Tokai Village, Japan (Fig.~\ref{fig:jparc}), and we are preparing for the experiment. We will use high purity pulsed proton beam from $3\ \mathrm{GeV}$ rapid cycling synchrotron (RCS) with operating fast extraction. The design power of the RCS is $1\ \mathrm{MW}$, and $500\ \mathrm{kW}$ operation has been achieved as of November, 2019. The pulsed proton beams have a $25\ \mathrm{Hz}$ double-pulsed structure.
\subsection{Sensitivity Goal}
\noindent The goal of our experiment is to achieve a single event sensitivity of $1\times10^{-13}$ for a graphite target or $2\times10^{-14}$ for a silicon carbide target for 1 year. The current limit for $\mu$-$e$ conversion is $4.6\times10^{-12}$ for a titanium target by the experiment at TRIUMF, and $4.3\times10^{-12}$ for a titanium target and $7\times10^{-13}$ for a gold target by the SINDRUM-II experiment. We therefore aim to improve the current limit by one or two orders or magnitude.
\subsection{Experimental Concept}
\noindent In the Standard Model, after a muon stopping in a nucleus to form a muonic atom, the muon can decay in orbit, or be captured by the nucleus. The probabilities at which these processes occur depend on the nuclear mass, for example, in a carbon muonic atom, 92\% will be decay-in-orbit, and 8\% caputured, while 33\% decay-in-orbit and 67\% captured in a silicon muonic atom.\\ 
\indent We will look for a signal of $\mu$-$e$ conversion $\mu^{-}$N$ \rightarrow e^{-}$N (N is a nucleus). The conversion electron is monoenergetic, with an energy equal to the muon mass minus the binding energy of the muon, and the nuclear recoil. For a light muonic atom such as carbon or silicon that will be used in the experiment, the experimental signature will be an electron with an energy of approximately $105\ \mathrm{MeV}$.\\
\indent Figure \ref{fig:hline2} shows the experimental setup in G4beamline simulation. First, pulsed proton beams are injected into the target made of carbon to produce pions. The pions then decay-in-flight into muons, and the target atoms and muons form muonic atoms. In the experiment, we will use only one target for pion production, pion decay, and muon stopping.\\
\indent After extracting the decaying particles through the secondary beamline, H Line, and removing the low-momentum backgrounds, we will search for the signal by using a magnetic spectrometer, which consists of a spectrometer electromagnet and four tracking detectors (Fig. \ref{fig:hline1}).
\begin{center}
\begin{figure}[htbp]
\begin{minipage}[t]{0.09\hsize}
\hspace{0.09\hsize}
\end{minipage}
\begin{minipage}[t]{0.4\hsize}
  \centering
  \includegraphics[width=\hsize]{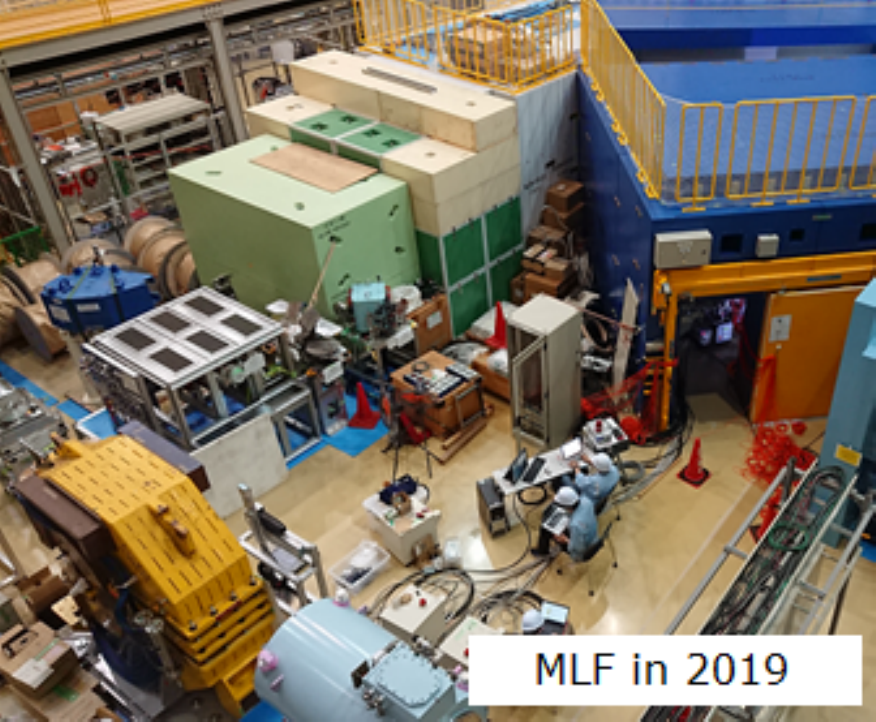}
  \caption{Photo of the MLF Experimental Hall No.1.}
  \label{fig:mlfphoto}
\end{minipage}
\begin{minipage}[t]{0.1\hsize}
\hspace{0.1\hsize}
\end{minipage}
\begin{minipage}[t]{0.28\hsize}
  \centering
  \includegraphics[width=\hsize]{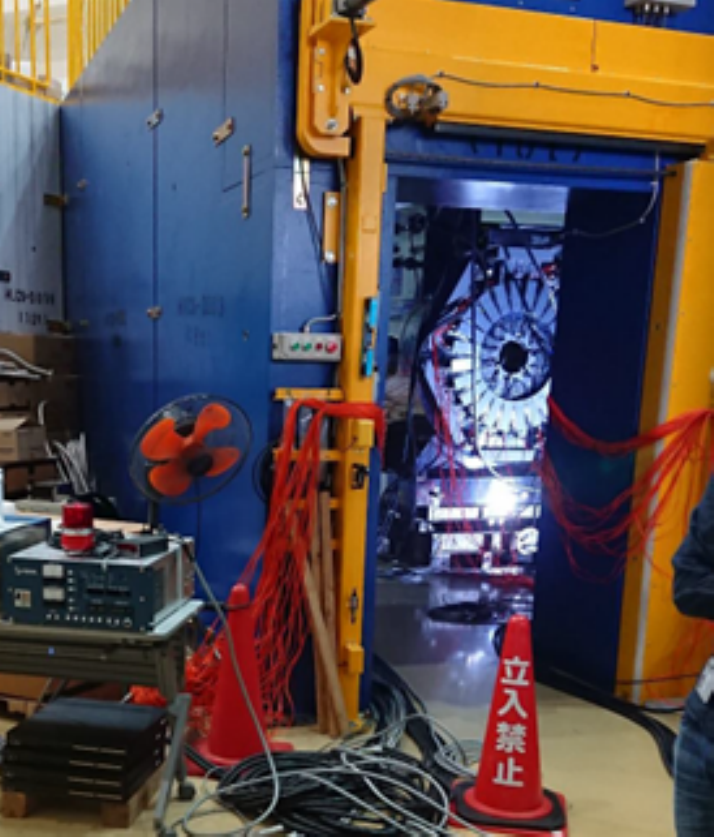}
  \caption{Photo inside the H1 Area.}
  \label{fig:hlinemag}
\end{minipage}
\end{figure}
\end{center}
\section{Current Status}
\subsection{The Secondary Beamline and H1 Area}
Figure \ref{fig:mlfphoto} shows a photo of the MLF Experimental Hall No.1. The radiation shield was built in 2016, and the construction of the H Line is ongoing (Fig. \ref{fig:hlinemag}). The magnetic spectrometer will be installed in the H1 Area.
\subsection{Spectrometer}
\noindent The dipole magnet of the spectrometer, which used to belong to the PIENU experiment in TRIUMF until 2012, was shipped to J-PARC in 2014. We tested the operation up to $500\ \mathrm{A}$, and measured the magnetic field. The magnetic field calculation using a simulator OPERA-3d agrees well with the measurement, and we will use the field map obtained from the calculation for tracking.\\
\indent Four tracking detectors, multi-wire proportional chambers (MWPCs) have been manufactured in 2015--2017 \cite{mwpc}, and the optimization study of the filling gas is in progress \cite{tau2018}. The amplifiers for all channels have been mass-produced. The FADCs of real-time lossless compression of waveform were developed \cite{fadc}. All the MWPCs and measuring equipment work well, and they are being used in the pre-experiment to measure the momentum of background described in the next section.
\subsection{Background Measurement at D2 Area}
\begin{figure}[htbp]
\begin{minipage}[t]{0.3\hsize}
\hspace{0.3\hsize}
\end{minipage}
\begin{minipage}[t]{0.4\hsize}
  \centering
  \includegraphics[width=\hsize]{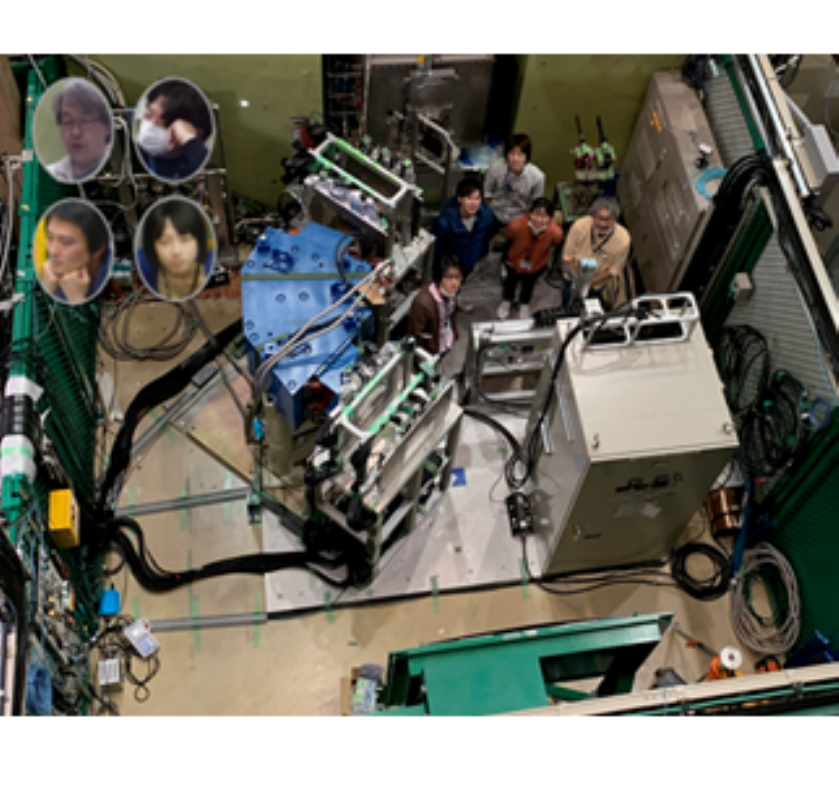}
  \caption{Experimental setup at D2 Area in March, 2019.}
  \label{fig:d2expphoto}
  \vspace{15pt}
\end{minipage}
\begin{minipage}[t]{0.23\hsize}
\hspace{0.23\hsize}
\end{minipage}
\end{figure}

\noindent We measured the momentum spectrum of electrons from muon decay in orbit $\mu^{-}$N$\rightarrow e^{-}\nu_{\mu}\overline{\nu_{e}}$N (muon DIO) around 45 to 55 $\mathrm{MeV}/c$ at D2 Area, MLF in March, 2019 \cite{nufact2018}. Pulsed muon beams hit a target made of carbon, and the momenta of decaying particles were measured by using a spectrometer, which consists of four MWPCs of final design and a magnet of KEK IMSS (Fig. \ref{fig:d2expphoto}).\\
\indent We have made the measurement three times in 2017 and 2019. Compared to the experiment in 2017, the tracking efficiency of the spectrometer improves by 1.6, and the time of data taking is extended. Table \ref{tb:exp} gives the summary of tracking events taken in the two experiments. Detailed analysis is in progress.
\begin{table}[htbp]
\caption{Summary and comparison of tracking events in the experiments in June, 2017 and March, 2019.}
\label{tb:exp}
\vspace{7pt}
\centering
\begin{tabular}{cccc} \hline
Mode & June, 2017 & March, 2019 & Ratio\\ \hline \hline
Muon DIO ($\mu^{-}$, $55\ \mathrm{MeV}/c$)  & 5028 events (63 runs) & 43769 events (346 runs) & 8.7\\ \hline
Michel edge ($\mu^{+}$, $52.5\ \mathrm{MeV}/c$)  & 23 runs & 35306 events (124 runs) & \\ \hline
Acceptance ($\mu^{+}$, $45\ \mathrm{MeV}/c$) & 16062 events (46 runs) & 86567 events (151 runs) & 5.4 \\ \hline
\vspace{15pt}
\end{tabular}
\end{table}
\section{Conclusion}
\noindent DeeMe is an experiment to search for $\mu$-$e$ conversion with a single event sensitivity of $10^{-14}$ using the carbon (or a silicon carbide) target. The construction of the secondary beamline, H Line, is in progress at J-PARC MLF. The spectrometer is ready, and we are optimizing the filling gas of the MWPCs for better performance.\\
\indent We measured the momentum of electrons from muon DIO using a carbon target around 45 to 55 $\mathrm{MeV}/c$ at MLF D2 Area in March, 2019. Detailed analysis is ongoing.

\end{document}